\documentstyle[aas2pp4]{article}

\def\'#1{\ifx#1i{\accent"13\i}\else{\accent"13#1}\fi}
\def\B{{\bf B}}
\def\cs{c_{\rm s}}
\def\gamef{\gamma_{\rm eff}}
\def\o{{\rm o}}

\def\r{{\bf r}}

\def\u{{\bf u}}
\def\urms{u_{\rm rms}}
\def\VS{V\'azquez-Semadeni}

\slugcomment{Submitted to {\it The Astrophysical Journal}}

\begin{document}

\title{Does Turbulent Pressure Behave as a Logatrope?}

   \author{Enrique V\'azquez-Semadeni, Jorge Cant\'o and Susana Lizano}
    \affil{Instituto de Astronom\'\i a, UNAM, Apdo. Postal 70-264, M\'exico, D.\
F.\ 04510, M\'exico}
\affil{e-mail: {\tt enro@astroscu.unam.mx}}

\begin{abstract}
We present numerical simulations of an isothermal turbulent gas undergoing
gravitational collapse, aimed at testing for ``logatropic''
behavior of the form $P_t \sim \log \rho$, where $P_t$ is the
``turbulent pressure'' and $\rho$ is the density. 
To this end, we monitor the evolution of the
turbulent velocity dispersion $\sigma$ as the density 
increases during the collapse. A logatropic behavior would require
that $\sigma \propto \rho^{-1/2}$, a result which, however, is not verified in
the simulations. Instead, the velocity dispersion {\it increases} with
density, implying a polytropic behavior of $P_t$. This behavior is found
both in purely hydrodynamic as well as hydromagnetic runs. For purely
hydrodynamic and rapidly-collapsing magnetic cases, 
the velocity dispersion increases roughly as
$\sigma \propto \rho^{1/2}$, implying $P_t\sim \rho^2$, where
$P_t$ is the turbulent pressure. For slowly-collapsing magnetic cases the
behavior is close to $\sigma \propto \rho^{1/4}$, which implies $P_t
\sim \rho^{3/2}$. We thus suggest that the
logatropic ``equation of state'' may represent only the statistically most
probable state of an ensemble of clouds in equilibrium between self-gravity and
kinetic support, but does not adequately
represent the behavior of the ``turbulent pressure'' within a cloud
undergoing a dynamic compression due to
gravitational collapse. Finally, we discuss the importance of the underlying
physical model for the clouds (in equilibrium vs.\ dynamic) 
on the results obtained.


\end{abstract}

Submitted to {\it The Astrophysical Journal}.

%
%________________________________________________________________

\section{Introduction}\label{introduction}

Molecular clouds and clumps exhibit the well-known velocity dispersion-
(or linewidth-)size relation 
\begin{equation}\label{larvel}
\sigma \sim R^{1/2},
\end{equation}
where $\sigma$ is the linewidth-determined velocity dispersion, and
$R$ the characteristic size. This correlation is observed both in
ensembles of clouds (Larson 1981; Leung et al.\ 1982;
Torrelles et al.\ 1983; Dame 
et al.\ 1986; Myers \& Goodman 1988; Falgarone, et al.\ 1992; Miesch
\& Bally 1994) or as a function of radius in quiescent cores using
various tracers (Fuller \& Myers 1992; Caselli \& Myers 1995;
Goodman et al.\ 1997), although the latter studies have suggested that
the scaling exponent in relation (\ref{larvel}) may actually differ between
massive and low-mass cores. Furthermore, Goodman et al.\ have
suggested that the exponent may decrease and approach zero as the
innermost regions of the cores are considered, in which the turbulent velocity
dispersion becomes subsonic.

A second scaling relation, between mean density $\langle\rho\rangle$
and size, is also generally reported, reading
\begin{equation}\label{lardens}
\langle\rho\rangle \sim R^{-1},
\end{equation}
although its authenticity has been questioned on theoretical (Kegel
1989; Scalo 1990) and numerical (V\'azquez-Semadeni,
Ballesteros-Paredes \& Rodr\'iguez 1997)
grounds, and significantly discrepant scaling exponents have been found
(e.g., Carr 1987; Loren 1989), or none at all (e.g., Plume et al.\
1997). Relations (\ref{larvel}) and (\ref{lardens}) constitute the
now famous ``Larson's relations''.

In spite of the anomalies at small scales and high-mass regions,
Larson's relations are generally accepted as
distinctive signatures of turbulence in molecular clouds and clumps
(e.g., Larson 1981; Scalo 1987), and together they imply
\begin{equation}\label{virial}
\sigma \propto \rho^{-1/2}.
\end{equation}
Relation (\ref{virial}) is actually a manifestation of virial
equilibrium between the turbulent velocity dispersion (possibly
magnetohydrodynamic, or MHD) in the clouds (Larson 1981).
A turbulent ``pressure'' $P_t$ corresponding to the turbulent velocity
dispersion can then be defined by (Lizano \& Shu 1989; hereafter LS)
\begin{equation}\label{turbpres}
\sigma^2 \equiv (dP_t/d\rho),
\end{equation}
in analogy with the relation between thermal pressure and the sound
speed. Choosing
\begin{equation}\label{logatrope}
P_t\propto \log \rho
\end{equation}
recovers the virial relation (\ref{virial}) (LS). 
Relation (\ref{logatrope}) is commonly refereed to as a
``logatropic equation of state''
%sls No entiendo el comentario, que acaso la ecuacion adiabatica
% $P=K \rho^\gamma$ no es una ecuacion de estado?
\footnote{Strictly speaking, this is
  not an equation of state, since it does not invlove all three
  thermodynamic variables. However, we will allow ourselves the
  terminology for consistency with common nomenclature.},
 or simply, a ``logatrope''.

%POSTSHORE
It is important to emphasize that the concept of a turbulent
``pressure'' may not be a very realistic representation of the effects of
turbulence, since it implicitly assumes a microscopic and
isotropic process. Instead, turbulence is a phenomenon involving a
wide range of 
spatial scales, from the scale size of the system under consideration
to the smallest dissipative scales. In particular, the existence of
large-scale modes implies
coherent motions which are more akin to ram pressure (locally
anisotropic, with a
well-defined direction) than to an isotropic, thermodynamic 
pressure. For these reasons, in the present 
paper we will focus primarily on the turbulent velocity dispersion.
References to the turbulent ``pressure'' will be made assuming that
it can be defined according to eq.\ (\ref{turbpres}),
for compatibility with published work, but the above
caveat should always be kept in mind.
%END POSTSHORE

The logatropic equation of state has been used in a number of studies
of cloud support and stability, such as quasi-static contraction
(LS), nonlinear wave propagation (Adams \& Fatuzzo
1993; Adams, Fatuzzo \& Watkins 1994; Gehman et al.\ 1996), and
gravitational stability (McLaughlin \& Pudritz 1996), among others.
Nevertheless, the logatropic equation remains a completely empirical
assumption, and there is no direct evidence that turbulent pressure
indeed behaves in this manner in fully dynamic situations. 
In fact, there are some lines of
reasoning which suggest that it may not:

\noindent
1. Larson's relations are observed in either {\it ensembles} of
relaxed clouds (e.g., Torrelles et al.\ 1983; Dame
et al.\ 1986; Myers \& Goodman 1988; Falgarone, et al.\ 1992; Miesch
\& Bally 1994), or as a function of radius in {\it quiescent} cores 
(Fuller \& Myers 1992; Caselli \& Myers 1995;
Goodman et al.\ 1997), but there is no evidence that they hold 
in fully dynamical processes, such as gravitational collapse. 
Interestingly, clouds which are strongly perturbed also
seem to not follow the Larson scaling relations (e.g.,
Loren 1989; Plume et al.\ 1997). In general, there is
little sampling of fully out-of-equilibrium, dynamical
processes and, in particular, dynamical collapse has not yet been 
directly detected because
dynamical velocities occur only at very small scales. 

\noindent
2. Although it might be argued that an ensemble of molecular clouds provides a
complete sample of various dynamical stages, in actuality most
observations refer to clouds and clumps close to {\it
equilibrium}. 
Thus, the clouds included in surveys such as Larson's (1989)
constitute an ensemble of equilibrium states for clouds of different
masses rather than an ensemble of evolutionary steps for a single
cloud (of constant mass). That is, instead of representing a number of
different states for the same cloud, they represent the same state for
different clouds.

\noindent
3. V\'azquez-Semadeni et al.\ (1997) have suggested that there may exist
large numbers of low-column density clouds which do not
satisfy the density-size relation, and that possibly only the
highest-column density clouds follow such a scaling
relation. Thus, the logatropic equation of state is not expected to
apply to such low-column density clouds, which are probably not in
self-gravitating equilibrium, but rather pressure- or ram pressure-confined.
%sls  estos objetos son transientes?

As a first attempt to decide on this matter, in this paper we present
two-dimensional numerical simulations of a turbulent,
self-gravitating, magnetized, isothermal gas, aimed
at testing the variation of the velocity dispersion as a cloud is
compressed by self-gravity. A related calculation has been performed
by Bonazzola et al.\ (1987), who used low-resolution simulations to
estimate the correlation between the nonlinear advection
term (related to the turbulent pressure) and the density gradient in a
compressible turbulent flow. 

%POSTREF
We emphasize that the simulations discussed in
this paper are not presented as models of 
cloud cores and their observed linewidths, but only as numerical
``experiments'' designed to test the applicability of the logatropic
equation of state. Furthermore, throughout this paper we will
refer exclusively to the {\it non-thermal} part of the velocity
dispersion. In contrast with the observational situation, where the
separation between the thermal and non-thermal components is an
issue (e.g., Fuller \& Myers 1992), in the simulations
this is a trivial 
task, since there is no confusion between the fluid velocity and
the thermal velocity dispersion, the latter being directly represented by the
temperature field.
%END POSTREF

The outline of the present paper is as
follows. In \S \ref{model} we describe the
numerical model; in \S \ref{results} we present the results, for both
purely hydrodynamic and fully MHD cases, and in \S
\ref{conclusions} we summarize and discuss our results.

\section{Numerical model}\label{model}
We numerically solve the full MHD equations in two or three dimensions (2D and
3D, respectively) in the presence of
self-gravity for an isothermal, grav\-i\-ta\-tion\-al\-ly-unstable gas, using
the pseudo-spectral code described in \VS, Passot \& Pouquet (1996),
although here we restrict ourselves to a scale-free, isothermal case. The
equations read
\begin{equation}
{\partial\rho\over\partial t} + {\nabla}\cdot (\rho\u) = \mu
\nabla^2 \rho, \label{cont}
\end{equation}
\begin{eqnarray}
{\partial\u\over\partial t} + \u\cdot\nabla\u =
-{\nabla P\over \rho}
- \Bigl({J \over M_a}\Bigr)^2 \nabla \phi +\nonumber \\
{1 \over \rho}
\bigl(\nabla \times {\bf B}\bigr) \times {\bf B}
 - {\nu_8} {\nabla^8\u}+\nonumber \\
\nu_2 (\nabla^2 \u + \frac{1}{3} \nabla \nabla \cdot \u),\label{mom}
\end{eqnarray}
\begin{equation}
{\partial {\bf B}\over\partial t} = \nabla \times (\u \times {\bf B})
- {\nu_8} {\nabla^8{\bf B}} + \eta \nabla^2 {\bf B}, \label{magn}
\end{equation}
\begin{equation}
\nabla^2 \phi=\rho -1, \label{poisson}
\end{equation}
and
\begin{equation}\label{eqstat}
P = c^2 \rho,
\end{equation}
where, as usual, $\rho$ is the density, {\u} is the fluid velocity,
$P$ is the thermal pressure, \B\ is the magnetic induction, and $\phi$
is the gravitational potential. The nondimensionalization is the
same as in Passot, V\'azquez-Semadeni \& Pouquet (1995), 
to which we refer the reader for
more details of the numerical method. The units are $\rho_o = \langle
\rho \rangle$ (mean
density in the integration domain), $u_\o = \cs$ (isothermal speed of
sound), $L_\o$ (size of the integration domain $=2 \pi$), $t_\o =  L_\o/u_\o$
(sound crossing time for the integration box), and $B_\o$ (magnetic field
strength such that at $\rho = \rho_\o$, $v_A = u_\o =\cs$, 
where $v_A$ is the Alfv\'en speed). The resulting nondimensional parameters
are $J \equiv L_\o/L_{\rm J}$, 
the number of Jeans lengths in the integration box and 
the Mach number of the velocity unit
$M_a = u_\o/\cs$. Due to our choice of units, $M_a\equiv 1$. 

Since the method is pseudospectral, it uses periodic boundary
conditions and has no numerical dissipation. Due to the latter,
dissipative operators need to be included explicitly. We use a
combination of $\nabla^8$ ``hyperviscosity'' and standard second-order
viscosity in the momentum and magnetic equations which allows
confinement of dissipative effects to the smallest scales in the
simulation while filtering oscillations in the vicinity of strong 
shocks (Passot \&
Pouquet 1988; see also the discussion by \VS ~ et al.\
1996). We also use a small amount of diffusion in the continuity
equation (\ref{cont}) which also helps the code to handle strong
shocks. Finally, the Poisson equation 
(\ref{poisson}) is used in a form suitable for
handling infinite or periodic media (Alecian \& L\'eorat 1988).

The initial conditions for the simulations have a smooth Gaussian
density profile peaked at the center of the integration box, with
$\rho_{\rm max}= 3.35 \rho_\o$ and a FWHM of $0.53 L_\o$. The initial
velocity field is turbulent with Gaussian fluctuations of rms
amplitude $\urms = 0.8 \cs$ and random phases, with exclusively
rotational modes (i.e., no compressible motions). 

We have performed three non-magnetic simulations in two dimensions 
at resolutions of
128, 256 and 512 grid points per dimension, respectively labeled R128,
R256 and R512. Except for the resolution and the amount of dissipation
(smaller at higher resolution), these runs are otherwise identical.
A non-turbulent run at resolution of 512 grid point per dimension
labeled NT512 was
also performed in order to test the numerical noise due to the
grid discreteness in the calculation of the velocity dispersion (see
below). Finally, a three-dimensional run, labeled run 3D96, 
with a resolution of 96 grid
points per dimension was also performed in order to test for dimensional
effects.

Additionally, we performed four magnetic simulations. 
For these runs, labeled M256, MM256, M512 and MM512, 
the initial magnetic field is along the $x$-direction, 
with strength $B_x=0.2$ for the ``M'' runs and $B_x=1.0$ for the ``MM'' runs. 
For all runs except MM256 and MM512, the Jeans length is $L_{\rm J}=0.9
L_\o$, while for MM256 and MM512, $L_{\rm J}=0.67 L_\o$. A summary of the runs
and their parameters is given in Table 1, including the diffusion coefficients.
In all cases, $\eta =0.002$.

The purpose of performing the three non-magnetic simulations at
different resolutions is to test for convergence, in particular with
regard to the effect of dissipation. Concerning the magnetic simulations,
runs M256 and M512 are
respectively similar to R256 and R512 except for the inclusion of the
magnetic field. However, the magnetic field strength used in the ``M'' runs is
quite small, so that the magnetic field does not prevent the
gravitational collapse. Runs MM256 and MM512 have a larger magnetic
field (implying an Alfv\'en speed equal to the sound
speed), somewhat closer to molecular cloud conditions. In
order to guarantee that the MM runs still undergo gravitational collapse, a
smaller Jeans length was used.

In the simulations we define the ``collapsing cloud'' as a circular region
within the simulation, centered at the peak of the density
distribution (calculated each time), containing 30\% of the total
mass. For the turbulent runs this may not be the best
approximation, since the cloud shape is not really circular (see fig.\
1). However, using a true Lagrangian cloud boundary would be much more
numerically involved, and we feel that our definition still provides
reasonably accurate results, since ultimately gravity overpowers the
turbulence and the shapes do not differ substantially from circular.
For the 3D run, the cloud was defined as the region containing 10\% of the
total mass.

The computation of the velocity dispersion requires some special care
in order to remove the bulk infall velocity. 
%POSTREF
This is a necessary step, since by definition the velocity dispersion
is the root mean square velocity fluctuation, i.e., $\langle (\u -
\langle \u \rangle)^2 \rangle$. However, in the present case of a
collapsing cloud, the mean velocity is a function of
radius.\footnote{In general, the mean velocity will always be a
  function of position in the case of a gas mass
  undergoing a global volume change. The simplest example is that of a
  gas mass in a cubic container being compressed by a piston on one
  side. In the direction of compression, the mean flow velocity will
  be a function of position, being zero at the fixed wall, and equal
  to the velocity of the piston at the side of the piston. If
  the gas is additionally turbulent, the turbulent motions will be
  superposed on top of this mean-flow velocity.} Thus, we use the
following procedure. 
%END POSTREF
We first compute the
average infall speed $u_r$ as a function of radius, 
and then compute the velocity dispersion as $\sigma \equiv
\langle\bigl(\u(\r)-u_r(r) {\bf \hat r}\bigr)^2\rangle^{1/2}$, where ${\bf \hat
r}$ is the unit vector in the radial direction, and the
%sls aqui todo esta sin dimensiones, verdad?
average is taken over the whole cloud, but using $r=|\r|$ at each
position. This procedure shows why our Cartesian grid introduces
noise: the ``circular'' paths along which $u_r$ is computed are not a perfect
circumference, but rather the best possible approximation to one on a
Cartesian grid, and the grid points on the path are not all at exactly
the same distance from the center. Thus, even in the non-turbulent
case there will be a systematic velocity dispersion at every radius,
due to the presence of a radial velocity gradient and to the
``thickness'' of the circumference, 
the error being largest at the smallest radii. In order to estimate
the magnitude of the numerical noise, we
computed the numerical velocity dispersion in the nonturbulent case as well.

\section{Results}\label{results}

As a typical example, figures 1 and 2 show
the density and velocity dispersion fields at times $t=0.9$ and
$t=2.4$ in non-dimensional units
for run R512, repectively. The former corresponds to the time
of minimum velocity dispersion, after shocks have dissipated the
initial velocity dispersion to a more slowly-decaying level, but
before compression has begun to enlarge it again (see discussion about fig.\
3 below).
The latter time is the final state of the simulation, after which 
it stops because the code cannot handle
the very large gradients developed at the center
of the cloud any more.

Figure 3 shows the log of the velocity dispersion vs.\ the log of the mean
density for all runs as they evolve. 
For all the turbulent cases, it is seen that the
initial transients suffer significant dissipation through shocks until
a less dissipative regime is reached, at which the velocity dispersion is at a
minimum. Subsequently, the velocity
dispersion tends to {\it increase} with mean density, although with
significant fluctuations. This behavior is in sharp contrast with
relation (\ref{virial}). For the purely hydrodynamic runs
R128, R256 and R512, a trend towards longer periods 
(i.e., larger density ranges) of nearly
power-law behavior is noticeable. Although
convergence on the duration of the power-law epoch of the evolution
may have not been reached yet at $512^2$
resolution, the slope does appear to be converging to a value of
1/2. Even if convergence has not been attained yet, the observed
trend is towards {\it steeper} slopes at higher resolutions, so in any
case the discrepancy with relation (\ref{virial}) appears robust.

An important possible problem is that this result might be an effect of the 
two-dimensionality of the simulations. Run 3D96 was performed as an attempt to
resolve this question, although the resolution is necessarily lower. 
The evolution of the velocity dispersion and the mean density
for this run is also shown in fig.\ 3.
Although at much slower rates than in the 2D runs due to the higher
dissipation inherent to the lower resolution, the trend in run 3D96 is still 
towards increasing $\sigma$ with $\langle \rho \rangle$ after the initial
transients have passed. Thus, even though run 3D96 does not permit confirmation
of the rates approached by the high-resolution 2D runs, the increasing trend of
$\sigma$ with $\langle \rho \rangle$ is maintained, suggesting that
this behavior is real, rather than just an effect of the
two-dimensionality. 

%POSTREF
In this regard, note that in general the simulations overestimate the
viscous dissipation rate, since, for
numerical reasons, the viscous coefficients have to be chosen so that
the dissipative scales fit within the resolution of the
simulation. Instead, in the actual interstellar gas, the dissipation
scales may be many orders of magnitude smaller than the scales of
interest. Thus, the dissipation rate of $\sigma$ found in the
simulations is at worst a lower bound to the actual rate, and the net
increase of $\sigma$ found in the simulations is expected to be a real
effect. In particular, run 3D96 is the most dissipative of the runs
performed, but a net increase is found also in this case.
%END POSTREF

The non-turbulent run NT512 also exhibits a velocity dispersion which
increases with mean density. This should be interpreted as a
numerically-generated velocity dispersion which increases at larger
infall speeds because the radial velocity gradient is also larger. 
Nevertheless, this numerical velocity dispersion is seen
to be generally about two orders of magnitude smaller than that for
the turbulent runs (note that the curve for the non-turbulent run has
been displaced upwards by an amount of 1.5 in $\log \sigma$ so that it fits
within the plot). Thus, we rule out numerical noise as the cause
for the trends observed for the turbulent runs in fig.\ 3.

The magnetic runs also exhibit a trend of increasing velocity
dispersion with increasing mean density, although with quite stronger
fluctuations. Runs MM256 and MM512 exhibit a range of
densities for which again roughly $\sigma \sim \rho^{1/2}$. Runs M256
and M512 exhibit a somewhat slower dependence, 
close to $\sigma \sim \rho^{1/4}$. In any case, the general
trend is the same as in the non-magnetic cases, contrary to the
logatropic behavior, relation (\ref{virial}).

\section{Conclusions}\label{conclusions}

\subsection{Summary and discussion}\label{sumandisc}

We have argued that Larson's (1981) relations and the resulting logatropic
``equation of state'' (relation [\ref{logatrope}]) 
and virial condition (relation [\ref{virial}]) may describe an 
ensemble of clouds {\it in
(near) equilibrium} between self-gravity and the turbulent velocity
dispersion, but not out-of-equilibrium, 
dynamical processes occurring on a single cloud.
We have tested this assertion by means of 
numerical simulations of collapsing clouds with an initial turbulent
velocity field, in both magnetic and non-magnetic regimes.

The simulations exhibit in all cases a turbulent
velocity dispersion which {\it increases} with mean density as the
collapse proceeds, in contradiction with the expected behavior for a
logatrope, relation (\ref{virial}). Non-magnetic and strongly self-gravitating
runs seem to
approach a power-law behavior of the form $\sigma \sim \langle \rho
\rangle^{1/2}$, while weakly self-gravitating 
magnetic runs in general tend to have shallower
dependences, although always with positive exponents. In particular,
the fact that magnetic runs exhibit the same qualitative behavior
suggests that weak magnetic fields cannot induce a logatropic (or  a
$\sigma \sim \rho^{-1/2)}$ behavior either. Interestingly, 
run M512 exhibits a behavior very close to $\sigma \sim
\rho^{1/4}$. Assuming that this run has converged 
to the true slope, it is noteworthy that
the implied turbulent pressure satisfies $P_t 
\sim \rho^{3/2}$. This result is consistent with that of McKee \& Zweibel
(1995) for the polytropic index of Alfv\'en waves under slow
compression. However, runs MM256 and MM512 appear to be closer
to the $\sigma \sim \rho^{1/2}$ ($P_t \propto \rho^2$)
behavior observed in the non-magnetic
runs. This distinction 
is likely to be due to the larger Jeans length used in the
M runs, implying a slower collapse 
(final time $t_{\rm fin}=2.4$) than for the MM runs ($t_{\rm fin}=1.0$), so
that the M runs are closer to the slow compression assumption of McKee and
Zweibel.

We emphasize that although convergence may not have been fully
achieved yet at the highest resolution we used ($512 \times 512$ grid
points), the trend is towards
{\it faster} increase of the velocity dispersion with density at
higher resolution, away from the behavior predicted by the
logatropic equation. Thus, the result that the velocity dispersion
increases with mean density appears quite robust.
Moreover, the trend of increasing $\sigma$ with $\langle \rho \rangle$ is
preserved in run 3D96 (albeit at a slower rate due to the lower resolution of
this run), thus ruling out the possibility that our results are a purely 2D
effect.

The main consequence of our results is that the logatropic ``equation
of state'' appears to be inadequate for the description of dynamical processes
occurring in a cloud.
%; specifically, for processes whose characteristic
%timescales are shorter or comparable to the characteristic
%virialization time within the cloud, in turn presumably comparable
%with the sound (or Alfv\'en, in the magnetic cases) crossing time for
%the cloud. 
This implies that the use of the logatropic equation in
studies of gravitational collapse and dynamical stability 
%(e.g., McLaughlin \& Pudritz 1996)
is questionable. Its use in studies
of quasi-static processes (e.g., LS) may still be
justified, although in general the question remains open as to whether
the logatropic equation, which can be thought of as representing the
{\it final} states of the virialization process, also represents the
behavior of the turbulent pressure {\it during} the relaxation processes
which lead to virialization. For this reason, it would also be interesting to
test its applicability in problems of nonlinear wave propagation 
(e.g.,Adams \& Fatuzzo 1993; Adams, Fatuzzo \& Watkins 1994; Gehman et al.\
1996).

Finally, we remark that the ensemble consisting of the evolutionary states of
our simulated gravitationally-collapsing clouds with a fixed mass is
completely different from the ensemble constructed from the observations of
many clouds of different masses in near equilibrium. The present work shows
that for the former ensemble, the logatropic
equation of state is not applicable.

\subsection{Comparison with previous work}\label{comparison}

In our simulations we obtain a polytropic form ($P_t \propto \rho^{\gamef}$)
for the effective ``equation of state'' of the turbulence, with
polytropic exponents $\gamef=2$ for the non-magnetic and strongly
self-gravitating cases, and $\gamef=3/2$ for the weakly self-gravitating
cases.
This result appears to be in contradiction with that of McLaughlin \& Pudritz
(1996, hereafter MP), who conclude that the total pressure 
(thermal plus turbulent) is
{\it not} expected to behave as a polytrope. MP reach this conclusion on the
basis of a stability analysis, noting that truncated 
polytropic solutions with $0 <
\gamef < 1$ (consistent with the observed lower temperatures of denser
structures) have never unstable, or even critically stable, equilibrium
solutions. That is, absolutely stable configurations are discarded by MP
so that a cloud is eventually able to collapse and form a star. 

It can then be seen that the difference between our results and those of MP
arises from the consideration of different physical models for the clouds. 
While MP's clouds are in hydrostatic equilibrium, our clouds 
are always out of equilibrium and are already unstable from the start. 
These may originate from clumps rendered unstable by
external turbulent compressions (V\'azquez-Semadeni et al.\ 1996)
if the effective equation of state (i.e., the heating
and cooling) permits it.
In such cases, the clumps never need to pass through a static equilibrium
state. Another possibility is the well-known onset of gravitational instability
due to the loss of magnetic support caused by ambipolar diffusion (e.g, Nakano
1979; LS).

Finally, we note that the polytropic exponents implied by our simulations are
larger than the critical value $\gamma_c=4/3$ below which gravitational
collapse can proceed to a singularity (e.g., Chandrasekhar 1961). Thus, if
turbulent pressure continued with this behavior unrestrictedly, it would
eventually halt the collapse. However, we do not expect this occur since, at
late stages of the collapse, dissipation becomes important again due to the
large velocity gradients that develop. In fact, in fig.\ 3 an end to
the steady increase of $\sigma$ is seen at large values of 
$\langle \rho \rangle$ for
several of the runs. Thus, we speculate that turbulent pressure cannot by
itself halt the collapse.

\begin{acknowledgements}

We would like to acknowledge John Scalo and Steve Shore for useful 
comments and discussions. This work has
made use of NASA's Astrophysical Data System Abstract Service.
The runs at $512 \times 512$ resolution and run 3D96 were performed on the
Cray YPM 4/64 of DGSCA, UNAM. This research has received partial financial
support from grants UNAM/DGAPA IN105295, UNAM/CRAY SC007196 and
CONACYT 4916-E9406.

\end{acknowledgements}

\clearpage
\vfill\eject

\begin{tabular}{lccllllll}
\hline\noalign{\smallskip}
Run & $\rho_{{\rm max}}/\langle\rho\rangle^{(a)}$ & FWHM/$L_o^{(b)}$ & 
\multicolumn{1}{c}{$L_J/L_o^{(c)}$} & 
\multicolumn{1}{c}{$u_{{\rm rms}}/C_s^{(d)}$} & 
\multicolumn{1}{c}{$B_x^{(e)}$} & 
\multicolumn{1}{c}{$\mu^{(f)}$} &
\multicolumn{1}{c}{$\nu_2^{(g)}$} &
\multicolumn{1}{c}{$\nu_8^{(h)}$}  \\
\noalign{\smallskip}
\hline\noalign{\smallskip}
R128  & 3.35 & 0.53 & 0.9  & 0.8 & 0 & 0.03 & $2.00 \times 10^{-3}$  & $8.12 \times 10^{-12}$\\
R256  & 3.35 & 0.53 & 0.9  & 0.8 & 0 & 0.0075 & $5.00\times 10^{-4}$ & $3.13\times 10^{-14}$\\
R512  & 3.35 & 0.53 & 0.9  & 0.8 & 0 & 0.008 & $1.25 \times 10^{-4}$ & $2.00 \times 10^{-16}$\\
NT512 & 3.35 & 0.53 & 0.9  & 0   & 0 & 0.008 & $1.25 \times 10^{-4}$ & $2.00 \times 10^{-16}$\\
M256  & 3.35 & 0.53 & 0.9  & 0.8 & 0.2 & 0.0075 & $5.00\times 10^{-4}$ & $3.13\times 10^{-14}$\\
M512  & 3.35 & 0.53 & 0.9  & 0.8 & 0.2 & 0.008 & $1.25 \times 10^{-4}$ & $2.00 \times 10^{-16}$\\
MM256 & 3.35 & 0.53 & 0.67 & 0.8 & 1.0 & 0.0075 & $5.00\times 10^{-4}$ & $3.13\times 10^{-14}$\\
MM512 & 3.35 & 0.53 & 0.67 & 0.8 & 1.0 & 0.008 & $1.25 \times 10^{-4}$ & $2.00 \times 10^{-16}$\\
3D96  & 4.93 & 0.54 & 0.9  & 0.8 & 0 & 0.013 & $1.30 \times 10^{-3}$ & $4.00\times 10^{-9}$\\
\noalign{\smallskip}
\hline\noalign{\smallskip}
\end{tabular}

\bigskip
\noindent
(a)  Central density of initial density peak in units of mean density \\
(b)  FWHM of initial density peak in units of box size \\
(c)  Jeans length in units of box size \\
(d)  Initial runs turbulent speed in units of sound speed \\
(e)  Initial strength of uniform magnetic field \\
(f)  Diffusion coefficient for the continuity equation\\
(g)  Standard viscosity coefficient\\
(h)  Hyperviscosity coefficient\\

\clearpage
\vfill\eject

\begin{figure}\label{collfig1}
\plotone{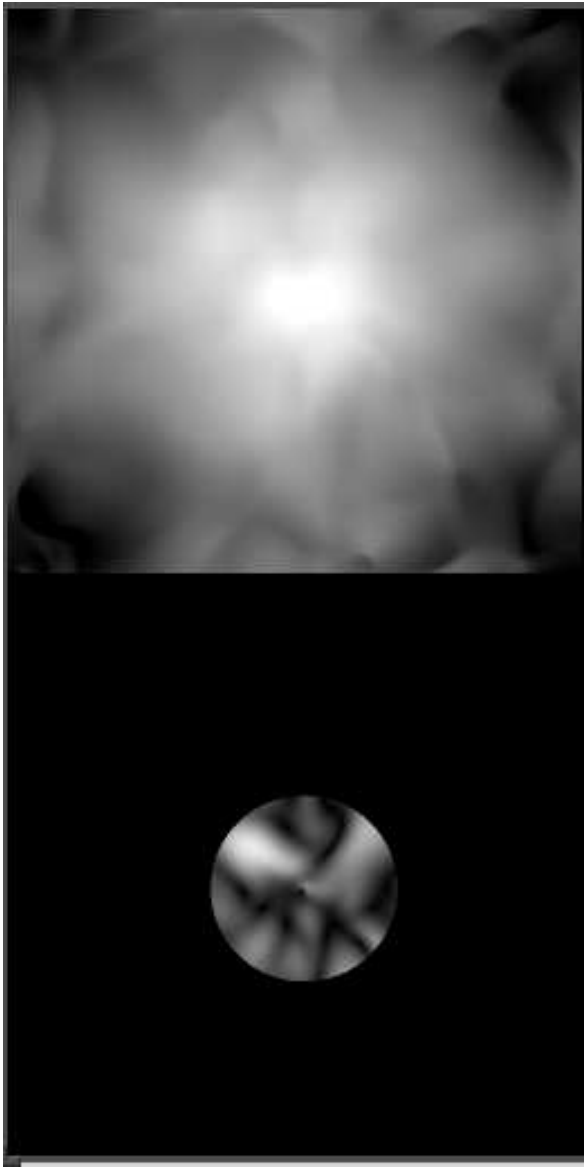}
\caption{Views of the density ({\it top}) and velocity dispersion
  ({\it bottom}) fields for run R512 at time $t=0.9$ in
  non-dimensional code units. This is the time
  of minimum velocity dispersion, having been already
  dissipated by shocks and not yet enhanced by the gravitational
  compression. The velocity dispersion is shown within the circle
  contining 0.3 of the total mass in the simulation. The gray scale
  for the density is logarithmic.}
\end{figure}
 
\begin{figure}\label{collfig2}
\plotone{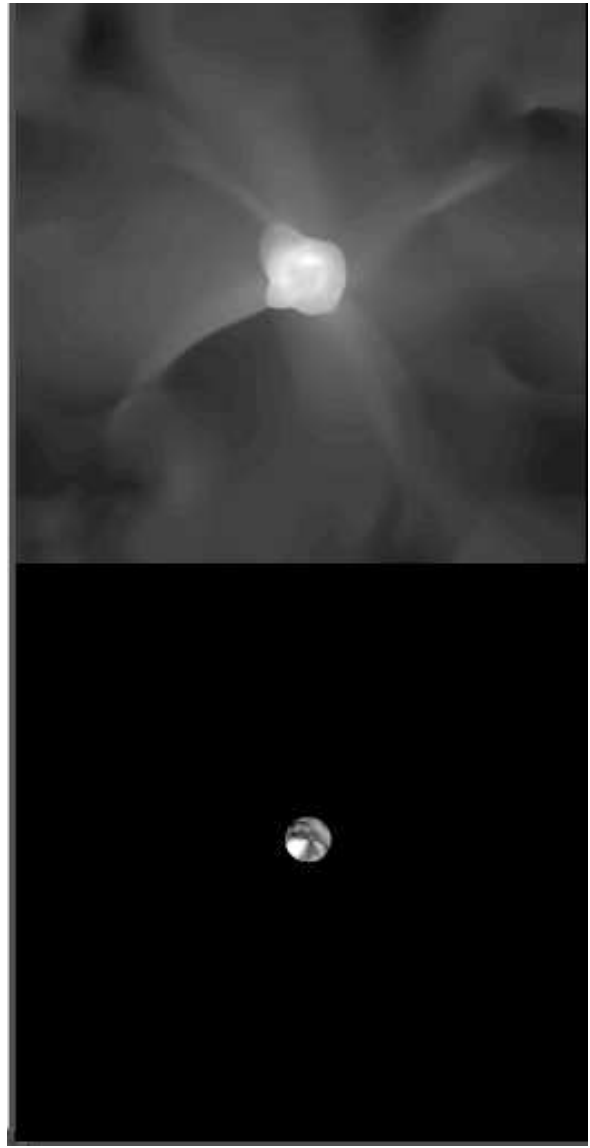}
\caption{Same as fig.\ 1 at time $t=2.4$. This is the
  final stage of the collapse. Note that the gray scale in this figure
  differs (has a larger maximum value) from that in
  fig.\ 1 in order to maximize clarity. The maximum density is $\rho_{\rm
    max} = 330 \langle \rho \rangle$.}
\end{figure}
 
\begin{figure}\label{sigma_den}
\plotone{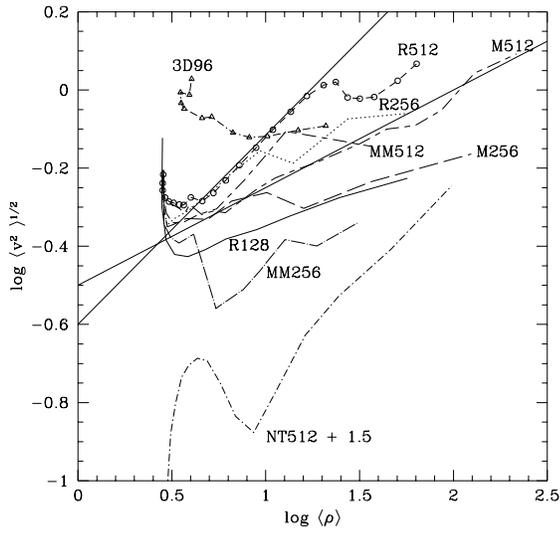}
\caption{Log-log plot of the velocity dispersion vs. average cloud
  density for all runs. The plot for run NT512 has been displaced
  upwards by an amount of 1.5 in the $y$-axis units. In all cases, the
  velocity dispersion is seen to increase with average density. The
  circles in the curve for run R512 indicate the individual times
  during the collapse at which the curve was sampled, taken at
  intervals $\Delta t= 0.1$ code time units. The solid straight
  line shows a power-law with exponent 1/2, and the dotted straight
  line shows a slope of 1/4.}
\end{figure}

\end{document}